\documentclass[sensors,article,submit,pdftex,moreauthors]{Definitions/mdpi} 

\firstpage{1} 
\pubvolume{1}
\issuenum{1}
\articlenumber{0}
\pubyear{2023}
\copyrightyear{2023}
\datereceived{ } 
\daterevised{ } 
\dateaccepted{ } 
\datepublished{ } 
\hreflink{https://doi.org/} 

\usepackage{siunitx}
\DeclareSIUnit\sqrthz{\ensuremath{\sqrt{\text{Hz}}}}
\DeclareSIUnit[per-mode = symbol]\masd{\m\per\sqrthz}

\usepackage{subcaption}

\Title{Nonlinearities in Long-Range Compact Michelson Interferometers}

\TitleCitation{Nonlinearities in Long-Range Compact Michelson Interferometers}

\Author{Jiri Smetana $^{1}$*\orcidA{}, Chiara Di Fronzo $^{2}$\orcidB{}, Anthony Amorosi $^{2}$\orcidC{} and Denis Martynov $^{1}$\orcidD{}}

\AuthorNames{Jiri Smetana, Chiara Di Fronzo, Anthony Amorosi and Denis Martynov}

\AuthorCitation{Smetana, J.; Di Fronzo, C.; Amorosi, A.; Martynov, D.}

\address{%
$^{1}$ \quad Institute for Gravitational Wave Astronomy, School of Physics and Astronomy, University of Birmingham, Birmingham B15 2TT, United Kingdom\\
$^{2}$ \quad Precision Mechatronics Laboratory, A\&M Dept., Universit\'e de Li\`ege, All\'e de la d\'ecouverte 9 B52/Quartier Polytec 1, B-4000 Li\`ege, Belgium}

\corres{Correspondence: gsmetana@star.sr.bham.ac.uk}

\abstract{Compact Michelson interferometers are well positioned to replace existing displacement sensors in the readout of seismometers and suspension systems, such as those used in contemporary gravitational-wave detectors. Here, we continue our previous investigation of a customised compact displacement sensor built by SmarAct, which operated on the principle of deep frequency modulation. The focus of this paper is on the linearity of this device. We show the three primary sources of nonlinearity that arise in the sensor---residual ellipticity, intrinsic distortion of the Lissajous figure, and distortion caused by exceeding the velocity limit imposed by the demodulation algorithm. We verify the theoretical models through an experimental demonstration designed to maximise the nonlinear noise to dominate regions of the readout's power spectrum. We finally simulate the effect that these nonlinearities are likely to have if implemented in the readout of the Advanced LIGO suspensions and show that the noise nonlinearities should not dominate across the key sub-\SI{10}{\Hz} frequency band.}

\keyword{displacement sensor; Michelson interferometer; nonlinearity} 

\preto{\abstractkeywords}{\nolinenumbers}

\begin{document}

\section{Introduction}
\label{sec:intro}

Interferometry sits at the forefront of high-precision displacement measurement. This technique sees use in a wide range of contexts concerned with the detection of weak signals, which originate from the minute scale of quantum mechanics through to the grand scale of astrophysical phenomena. Amongst the most notable interferometric devices are the contemporary gravitational-wave (GW) detectors, Advanced LIGO~\cite{AdvLIGO} and Advanced Virgo~\cite{AdvVirgo}, which are capable of \SI{2e-20}{\masd} precision in the peak sensitivity band around \SI{100}{\Hz}~\cite{Buikema2020}.

Although specialised detector facilities, such as GW detectors, can span kilometre scales, compact interferometric devices on the centimetre-scale can be traced back to 1972~\cite{Compact1972}. Since then, the sensitivity of such devices has continue to improve, with the LISA Pathfinder~\cite{CompactLISAPF} mission showing the versatility of interferometry in space-based applications. Advances in miniaturisation of optical and laser components have led to a range of compact devices that offer excellent sensitivity, as we outline below. The key to the interferometer's impressive sensitivity is in its extremely sharp response to small displacements, with the sensor's full range swept on the scale of the wavelength of light used. For example, the well-studied, sinusoidal response of a Michelson interferometer (e.g.~\cite{InterfTechniques}) operated with a typical \SI{1064}{\nm} laser covers the full signal range over a narrow span of just \SI{255}{\nm}.

A thorough overview of notable interferometric sensors can be found in Ref.~\cite{CompactReview}. In this paper we will focus principally on the performance of our custom-designed compact Michelson-type sensor built by the company SmarAct and analysed previously in Ref.~\cite{Smaract}. The nominal sub-picometre sensitivity that was achieved is useful across a range of applications (see Ref.~\cite{Smaract} and references therein), but we particularly focus on its utility in the sensing of quiet suspension systems, namely the quadruple suspension~\cite{LIGOQUAD} of the Advanced LIGO detectors. The current suspensions are sensed with BOSEM shadow sensors~\cite{Bosem}, which utilise an optical sensing scheme, albeit not an interferometric one. As argued in detail in Ref.~\cite{5Hz}, the low-frequency (5-30 Hz) band is limited by the injection of noise from angular control loops (also shown in Ref.~\cite{Buikema2020}), which can ultimately be traced back to the limiting sensitivity of the existing shadow sensors. Sensitivity improvements in this detection band are essential for enhancing early-warning systems~\cite{EarlyWarning,MultiMessenger} and expanding the range of detected GW sources towards intermediate-mass black holes~\cite{IMBHDetection}.

Existing GW detectors use a range of inertial and displacement sensors to improve the stability, provide active isolation of and provide readout for control of their suspension systems. However, the detectors stand to benefit from further improved inertial sensors. A notable way to improve these devices is with better sensors, with current devices being broadly limited at low frequencies by their readout noise. Interferometric sensors are well poised to address this problem, with one example, the HoQI~\cite{HoQITest}, already demonstrating an improvement in the low-frequency sensitivity of a commercial geophone~\cite{HoQIGeophone}. Recent high-precision devices, such as the BRS~\cite{BRS_2014,BRS} rotation sensor and 6D~\cite{6D,New6D} six-degree-of-freedom inertial sensor use interferometric sensing to achieve their advanced sensitivity. We expect that our sensor will be used in the future testing of the Compact-6D inertial sensor~\cite{C6D}---an evolution of the previous 6D design.

The path towards improved sensitivity in the key frequency band lies in the development of better sensors and the SmarAct interferometric sensor presents a good candidate for achieving this. However, to satisfy the requirements laid out in Ref.~\cite{5Hz} it will be important to reach the full sensitivity level demonstrated in Ref.~\cite{Smaract}. This is only possible if the performance of the sensor is not degraded once placed into the real environment rather than the typical `null measurement' setup used in its noise characterisation. Of particular interest to us is the impact of the sensor's linearity on the realistic sensitivity limit.

In this paper, we investigate the impact of nonlinear noise in Michelson-type interferometric displacement sensors when placed in high-RMS-displacement applications. The range of a simple Michelson interferometer is already narrow and its sinusoidal response means that the usable linear region yields an even smaller operating range. A number of different techniques exist in the literature~\cite{CompactReview} for extending this range through the use of multiple phase-offset readout channels, which allow for a linearised estimate of the displacement. Our particular readout scheme, based around the principle of deep frequency modulation (DFM)~\cite{DFM,DFM_2}, is discussed in more detail in Sect.~\ref{sec:model}. This readout scheme theoretically fully linearises the displacement and, with the use of a phase-unwrapping algorithm (known in this context as fringe counting)~\cite{FringeCount1,FringeCount2}, can extend the range of the interferometer over many multiples of the free spectral range (FSR). However, the linearisation algorithm can suffer from limitations that in practise lead to imperfect linearisation of the signal and the injection of periodic nonlinear error into the readout. The mechanisms of nonlinear coupling are analysed in Sec.~\ref{subsec:ellip}--\ref{subsec:demod}, with a theoretical framework laid out for modelling these nonlinearities in a real-world situation. The theoretical model is compared to measured data from an experimental scheme described in Sec~\ref{sec:experiment}. This theoretical model is finally applied to a simulation of the nonlinear noise performance in suspension sensing within the LIGO vacuum chambers in Sec.~\ref{sec:LIGO}.

\section{Modelling Nonlinearities}
\label{sec:model}

\begin{figure}[H]
    \begin{adjustwidth}{-\extralength}{0cm}
    \centering
    \begin{subfigure}{0.48\linewidth}
        \centering
        \caption{}
        \vspace{2mm}
        \includegraphics[width=\linewidth]{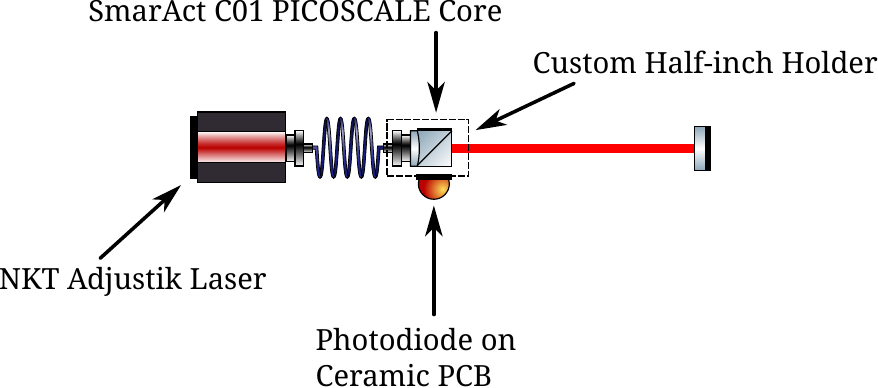}
        \label{subfig:sensor_schema}
    \end{subfigure}
    \hfill
    \begin{subfigure}{0.3\linewidth}
        \centering
        \caption{}
        \vspace{2mm}
        \includegraphics[width=\linewidth]{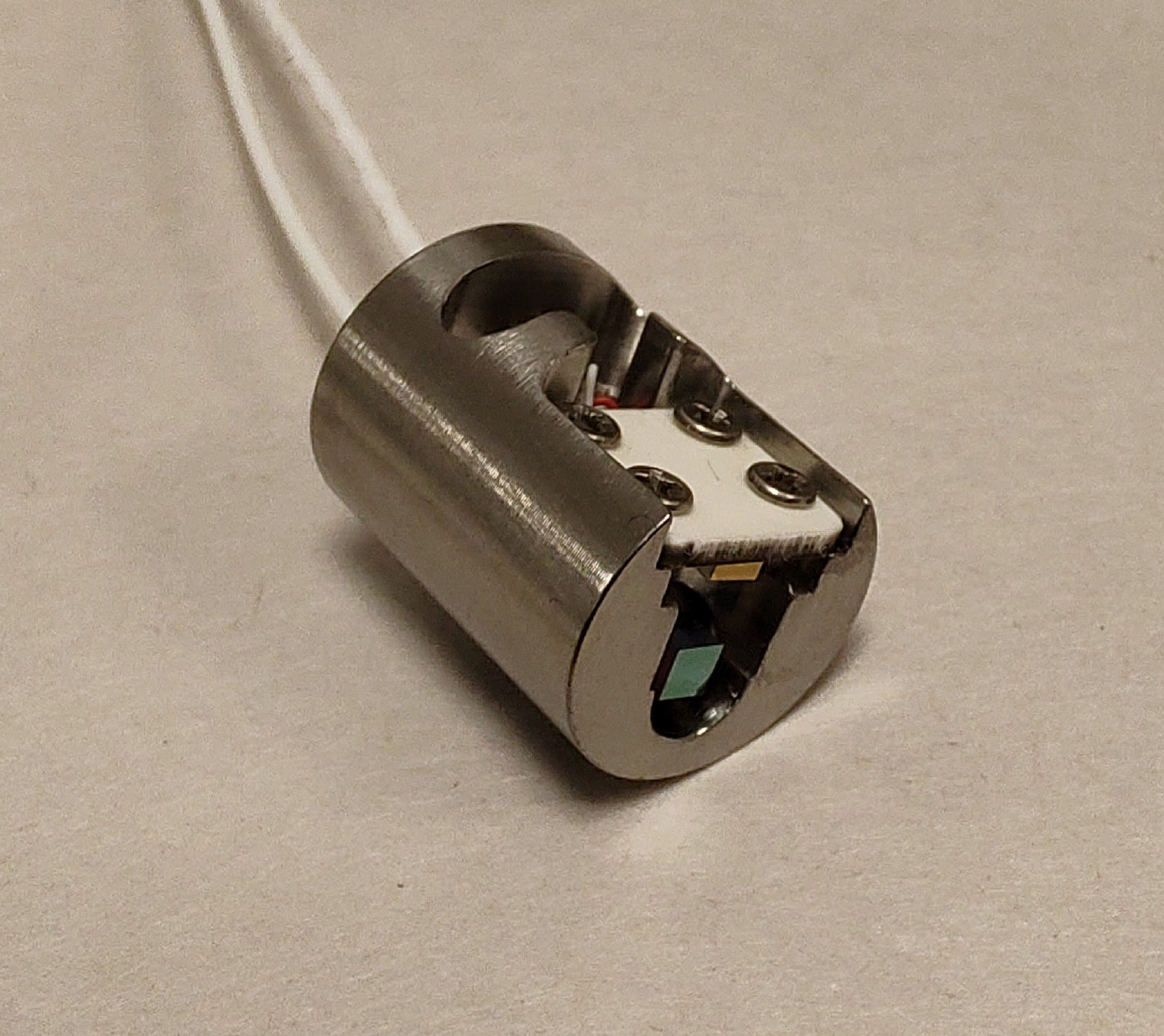}
        \label{subfig:sensor_photo}
    \end{subfigure}
    \end{adjustwidth}
    \caption{Schematic (a) and photo (b) of the custom-designed sensing head manufactured by SmarAct that is used throughout this work. The sensing head is embedded within the same readout structure as shown in Fig.~1 of Ref.~\cite{Smaract}.}
    \label{fig:sensor}
\end{figure}

The sensor scheme is based around a custom-designed opto-mechanical assembly derived from the SmarAct C01 PICOSCALE sensing head. The sensor, shown in Fig.~\ref{fig:sensor}, consists rather simply of a Michelson interferometer with an open port along one of the typical interferometer arms and a high-reflectivity coating applied to one face of the central beam splitter cube to act as the reference arm. This minimalist design leads to a highly compact sensor and mitigates losses from additional optical components. The simplicity of the design, a key aspect of the robust and sensitive scheme we investigated in Ref.~\cite{Smaract}, results in a comparatively greater level of complexity in the algorithm employed in the phase extraction scheme.

We make use of the DFM technique, described in depth in Refs.~\cite{DFM,DFM_2}, that sees growing applications in interferometric displacement measurement~\cite{Isleif_16,Gerberding_2017,Isleif_2019}. The scheme begins with a modulation of the laser frequency by any number of standard techniques, in our case control of the laser cavity piezoelectric transducer. Solving for the signal measured by a photodiode at the output of an unbalanced Michelson interferometer, we obtain the rather simple relation
\begin{equation}
    P(t) = A[1 + C \cos(\phi(t) + m \cos(\omega_m t))],
    \label{eq:power_raw}
\end{equation}
where $A$ is the signal scale factor (akin to an amplitude), $C$ is the fringe contrast, a value in the range 0--1 corresponding to the level of mode matching of the interfering beams, $\phi$ is the additional microscopic arm phase accumulated within the measurement arm of the interferometer, $m$ is the modulation index, and $\omega_m$ is the modulation angular frequency. The arm phase can be straightforwardly related to the mirror displacement via $x = \phi \lambda / (4 \pi)$. The modulation index can be intuitively written as $m = 4 \pi A_m \Delta L / c$, where $A_m$ is the modulation depth in frequency, such that the time-dependent laser frequency can be written as $f(t) = f_0 + A_m \cos(\omega_m t)$, and $\Delta L$ is the length difference between the reference arm and measurement arm of the interferometer. This result is commonly processed further in the limit of $m \ll 1$ where the small-angle approximations are appropriate and lead to a host of widespread uses, such as in cavity locking schemes (e.g. Pound-Drever-Hall locking~\cite{PDHScheme}). It is beneficial to consider the signal obtained in Eq.~\ref{eq:power_raw} as a Fourier series decomposition in terms of the harmonics of $\omega_m$, which is given by
\begin{equation}
    P(t) = P_0 + \sum_{n=1}^\infty 2 C A J_n(m) \cos(\phi + n \pi / 2) \cos(n \omega_m t),
    \label{eq:power_series}
\end{equation}
where $J_n$ is the $n^\mathrm{th}$-order Bessel function of the first kind and $P_0 = A(1 + C J_0(m) \cos(\phi))$. Whilst we may recover the typical small-angle approximation by considering only the $n=1$ term, we operate the system beyond the small-$m$ limit, which necessarily extends the scheme to include the higher order terms. This is important for us, as it gives us access to multiple signals with a cyclical sinusoidal dependence on the arm phase, $\phi$. We can thus take multiple signals to construct a linear estimator of the phase.

Our particular scheme relies on a technique of demodulation at multiple harmonics of $\omega_m$. If we multiply the raw signal by $cos(k \omega t)$ for integer vales of $k$, and filter out the beat signals above DC with an appropriate lowpass filter, we can write the $k^\mathrm{th}$ demodulated harmonic as
\begin{equation}
    S_k = C A J_k(m) \cos(\phi + k \pi / 2).
    \label{eq:demod}
\end{equation}
Assuming we have a stable scheme such that $A$, $C$ and $m$ remain constant, the simplest way to proceed is to take a pair of $S_k$ signals of different parity, for example from the $k=1$ and $k=2$ demodulation, and construct an elliptical Lissajous figure where the angular coordinate of a point along the Lissajous curve at any given time corresponds to the arm phase, $\phi$. This treatment simply lays out the key set of equations that are essential for understanding the origins of the nonlinear couplings that we discuss in following sections. Further details of the detection scheme, particularly the optical layout schematic and data acquisition process are laid out in our original work in Ref.~\cite{Smaract}.

\subsection{Nonlinear Effects on Sensitivity}
\label{subsec:sens}

Nonlinear effects in optical interferometry have been analysed in the past and are known to lead to a periodic error on the order of a few nanometres~\cite{Nonlin1,Nonlin2}. These effects commonly arise from cross-talk between the two nominally orthogonal signals. For example, due to imperfections in the polarisation optics in interferometers that utilise linearly polarised states of light such as the HoQI~\cite{HoQITest}.

In our case, the signals channels are well isolated from each other as their orthogonality is ensured by a rigid mathematical relation and not dependent on the quality/alignment of optical components. However, other types of common nonlinearities can still occur, for example the coupling of ghost beams into the readout port~\cite{GhostBeam, GhostBeamFibre,Gerberding_2021}. These nonlinear sources are important to consider but in our case are not the dominant effect due to the design of the sensor. Ghost beam effects are mitigated by the relatively few optical surfaces involved in the interferometric path. Furthermore, most ghost beams are generated through interactions between surfaces of the central beam cube, where drifts in the ghost beam phase arise due to thermal expansion of the beam cube. These effects are naturally mitigated by the cube's small volume (side length of 2 mm) and the good thermal properties of glass. Nonetheless, ghost beams can be problematic for high precision applications and their effects are being investigated. We can further improve the sensor's resilience to ghost beams by angling the input beam to the beam cube so avoid having any reflections off optical surfaces at normal inidence.

We find that the dominant source of nonlinearity arise from the particulars of the DFM scheme and can be traced back to the fidelity of the modulation-demodulation procedure, the accurate fitting of the parameters in Eq.~\ref{eq:demod}, and the limited bandwidth of the sensor. Following common wisdom, we should expect that nonlinearities increase disproportionately larger with growing signal RMS. The question is, however, whether these nonlinear effects can reduce the SNR below unity for a realistic RMS displacement or before reaching other natural constraints, such as the velocity limit imposed by the fringe-counting algorithm.

We consider a situation where high but realistic RMS displacement is reached due to a high-amplitude region of signal within a limited frequency band, with signal outside of this band settling at a much lower level by several orders of magnitude. Due to the nonlinear processes, it becomes possible for the high-amplitude signal to spread to other frequencies and thus swamp the true weak signals in the quiet regions of the spectrum. We investigate such a scenario, where the additional `nonlinear noise' generates a new and degraded noise performance, which prevents our device from reaching its nominal sub-picometre sensitivity.

This scenario is not entirely academic as it shares a practical similarity with the signals handled in inertial sensing devices. The response of a conventional mass-on-a-spring inertial sensor begins to substantially decrease towards DC, below the mass-spring resonant frequency. This feature can be found in commercial seismometers such as the Trillium T240, even more so in the velocity readout\footnote{Velocity readout multiplies the response to displacement in the frequency domain by an additional factor of $\omega$, thus further reducing the response towards DC.} of geophones such as the Sercel L-4C, but also in custom, precise angular sensors, such as the BRS and multi-DoF sensors such as 6D. In these applications, we are searching for very weak readout signals at low frequencies, whilst the high response at and above the resonance leads to a potentially large signal RMS.

\subsection{Nonlinearities from Ellipticity}
\label{subsec:ellip}

Our phase extraction algorithm takes multiple nonlinear functions of $\phi$ (specifically sinusoidal functions) and combines them to form a linearised readout. This algorithm relies on the correct knowledge or fitting of the ellipse parameters, which are used to circularise the ellipse for use with the four-quadrant arctangent function. Therefore, if there is a mismatch between the ellipse parameters, there will remain some residual ellipticity to the Lissajous figure, which will translate into a systematic nonlinear error.

We define a quantity, elliptical error, $\delta$, given by the fractional difference between the semi-major and semi-minor axes, such that the semi-major axis is given by $a = (1 + \delta) b$ for a semi-minor axis of size $b$. Through a Taylor expansion in $\delta$ (we can assume that in all reasonable scenarios $\delta \ll 1$), we find that the phase estimator, $\Bar{\phi}$, is related to the true arm phase approximately through
\begin{equation}
    \Bar{\phi} \approx \phi + \frac{\delta}{2} \sin(2 \phi),
\end{equation}
where the second term represents the first-order contribution to the periodic error that is generated as a result of the ellipticity.

We cannot proceed further in deriving a general spectral density equation for the nonlinear noise. However, we may use this as the basis of the time-domain simulations of the nonlinear effects that allow us to model and predict the nonlinear impact in particular applications. Additionally, due to the limited range of the sine function, we can make an estimate of the maximum noise floor that can be generated by this nonlinearity.

In the limit of broadband, high-RMS displacement, where the fluctuation in $\phi$ exceeds unity, we consider $\sin(2 \phi)$ to behave as a generator of random values within the interval [-1, 1]. Therefore, the variance of this term will be within a factor of a few below unity; we adopt a representative value of 1/3, corresponding to a uniform distribution. The power spectral density (PSD) that corresponds to this variance is broadened out as the original displacement spectrum saturates the sine function and spreads to other frequencies, leading to the approximate PSD of the sine term, $S_\textrm{sin} \approx 1 / 3\gamma_{eff}$, where $\gamma_{eff}$ is the effective bandwidth of the frequency-broadened signal. This bandwidth is highly variable based on the exact shape and RMS of the signal. However, as the noise is only ever broadened, we may state that the largest noise level that can be achieved is when $\gamma_{eff} = \omega_{hi}$, the upper edge of the frequency band of the original displacement signal. Thus our order-of-magnitude estimate for the maximum nonlinear noise PSD is given by
\begin{equation}
    S_x^\mathrm{max} = \frac{\lambda^2 \delta^2}{192 \pi^2 \omega_{hi}}.
\end{equation}
This equation only provides the maximal noise floor in the case of sufficiently broadband signal and is valid in the frequency band below $\gamma_{eff}$. As such, this noise level can be exceeded where nonlinear up-conversion occurs, particularly in the case of tightly localised resonances in the power spectrum that lead to the presence of prominent peaks at the higher harmonic frequencies.

This nonlinearity can arise due to a poor estimate of the ellipse parameters. To prevent this issue, it is possible to perform a slow sweep over the laser frequency, provided the laser frequency actuator has the range to sweep over at least one full FSR. We can fit to the resulting ellipse assuming the ellipse parameters do not change over time. In reality a number of effects can generate a drift in the ellipse parameters, such as nonlinearity in the modulation drive, timing jitter causing drifts in the demodulation phases, and residual amplitude modulation. We find that maintaining elliptical error at or below $\delta = 0.01$ over the long term is feasible.

\subsection{Nonlinearities from Nonellipticity}
\label{subsec:nonellip}

The extension of the above treatment of ellipticity leads us to consider nonlinearities that occur due to a departure from an elliptical Lissajous shape altogether. From our long-term observations, the sensor produces a high-fidelity elliptical Lissajous and, in most situations, the nonlinearities tend to be dominated by poorly fitted ellipse parameters. However, even with proper ellipse fitting in post processing, we find residual periodic error in the phase readout that suggests that the Lissajous figure in not entirely elliptical.

This departure of the Lissajous from a simple elliptical shape was already observed in the previous work in Ref.~\cite{Smaract}, where we showed the nonlinear error in the displacement readout, using a long-range ($\sim$100 FSR) scan over the laser frequency using the temperature setpoint. We revisit this result here, with a closer look at the amplitude and period of the error.

\begin{figure}[H]
    \centering
    \includegraphics[width=\linewidth]{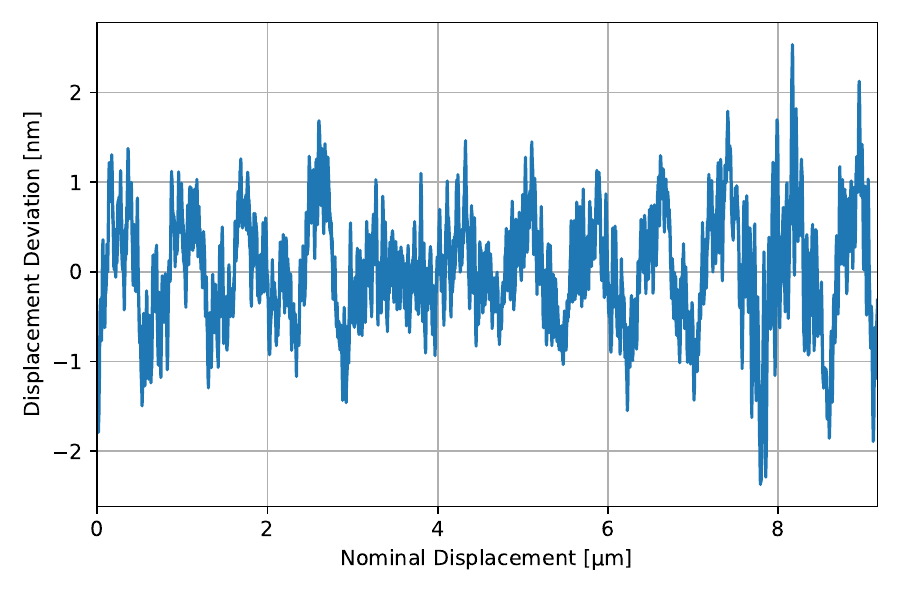}
    \caption{Nonlinear deviation of sensor readout from the `true' inferred displacement. The x-axis is calibrated into displacement by taking the displacement time series and fitting to the slow variations (not due to sensor nonlinearity) by a ninth-order polynomial.}
    \label{fig:nonellip}
\end{figure}

We sweep over the effective displacement by slowly actuating on the laser wavelength. A benefit of this method is that the modulation index does not change, which it would do if the sweep was performed over true displacement, thus introducing another source of nonlinearirty. This is achieved by setting the laser wavelength from one extreme to another (sweeping over a total wavelength span of \SI{0.9}{\nm}) and allowing the built-in temperature servo to shift the wavelength to the new setpoint. We isolate a region containing approximately 10 fringes in the middle of the sweep where the wavelength was swept through approximately linearly in time. We linearise the readout using an elliptical fit in post-processing and then remove the underlying, slowly varying, nonlinearity of the sweep using a ninth-order polynomial fit. As the period of the sensor nonlinearities is much shorter any trend in the sweep, we can remove these trends without also fitting to the periodic error. The result in Fig.~\ref{fig:nonellip}, shows the deviation of the measured displacement signal from the fitted `nominal' displacement over the swept region.

We note that the nonlinear deviation cannot be described by a simple function. However, it does show a clear periodicity that is approximately the same as that of the elliptical nonlinearity discussed in Sec.~\ref{subsec:ellip}. We can, therefore, assign an \textit{effective} elliptical error, $\delta_{eff}$, which should provide a estimate of the nonlinear impact to within a factor of a few. Taking the amplitude of the nonelliptical deviation to be \SI{1.2}{\nm}, this yields a $\delta_{eff}$ of around 2\%. As this nonlinearity arises through (as yet unknown) processes that distort the Lissajous figure away from an ideal ellipse, it is not clear how to suppress this nonlinearity further. Therefore, for now, this imposes a hard limit on the linearity of the sensor.

\subsection{Nonlinearities from Demodulation}
\label{subsec:demod}

The final nonlinearity we consider comes from the up-conversion of signal frequencies through the sine function interacting with the finite bandwidth of the sensor. Our algorithm relies on the low-passing of the demodulated signals. If we consider a purely sinusoidal displacement at some arbitrary frequency $\Omega$, leading to an arm-phase fluctuation given by $\phi = A \sin(\Omega t)$, our demodulated signals are proportional to $\sin(A \sin(\Omega t))$ (replacing the outer sine with cosine for the even harmonics).

This is a familiar equation within our setup and is of the type already encountered in Eq.~\ref{eq:power_raw}. Thus we can state that the demodulated signal can be written as an infinite series with terms proportional to $J_n(A) \sin(n \Omega t)$ for integer $n$. We can still reconstruct the signal to high enough accuracy only considering terms up to a particular order, $n$. To determine the maximum order that must be included, we make use of the fact that the $J_n(x)$ are diminishingly small for $x \ll x_m$, the location of their first maximum. We use the approximation that, for a particular order, the location of the first maximum is well approximated by the value of the order itself~\cite{Bessel}. From this we conclude that the critical order is given by $n \approx A$ and we can neglect all orders where $n \gg A$.

In a practical sense, this means that the maximum frequency that our original signal is significantly up-converted to is given by approximately $A\Omega$. This is consistent with the intuitive argument that the frequency of the outer sinusoid is given by the phase velocity, $\dot{\phi}$ and $A \Omega = \dot{\phi}_\mathrm{max}$. Taking this intuitive argument further, we propose that for any arbitrary displacement, the corresponding $\dot{\phi}_\mathrm{max}$ must be significantly below the cutoff frequency of the lowpass filter. In a sense the finite bandwidth of the sensor, $\gamma_{s}$, imposes a velocity limit on the sensor application, given by
\begin{equation}
    v_\mathrm{max} \ll \frac{\gamma_s \lambda}{4 \pi}.
\end{equation}

The first constraint on $\gamma_s$ is the lowpass filter cutoff frequency. However, this frequency cannot be increased arbitrarily, as when $\dot{\phi}_\mathrm{max}$ exceeds $\omega_m / 2$, the signal sidebands around the adjacent harmonics will leak into the measurement band of its neighbour harmonic and corrupt the signal there. Therefore, the sensor bandwidth is ultimately limited to $\gamma_s \leq \omega_m / 2$. Taking this limit for our setup, we obtain an absolute velocity limit of around \SI{50}{\um \per \s}.

\section{Sensitivity Degradation in a High-RMS Application}
\label{sec:experiment}

\begin{figure}[H]
    \begin{adjustwidth}{-\extralength}{0cm}
    \centering
    \begin{subfigure}{0.48\linewidth}
        \centering
        \caption{}
        \vspace{2mm}
        \includegraphics[width=\linewidth]{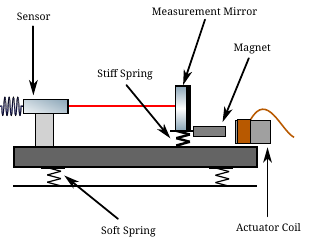}
        \label{subfig:layout_drawing}
    \end{subfigure}
    \hfill
    \begin{subfigure}{0.48\linewidth}
        \centering
        \caption{}
        \vspace{2mm}
        \includegraphics[width=\linewidth]{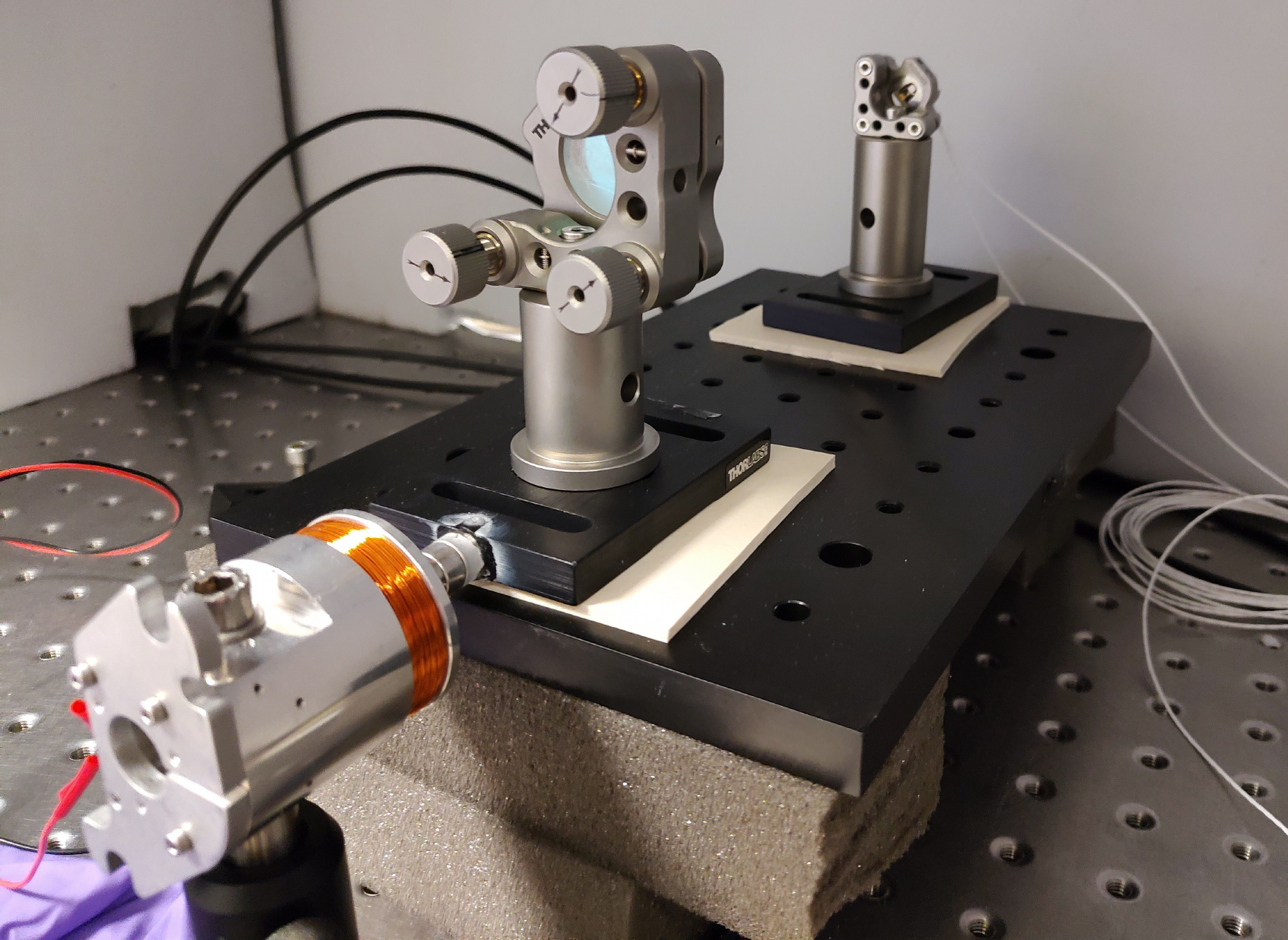}
        \label{subfig:layout_photo}
    \end{subfigure}
    \end{adjustwidth}
    \caption{Adjusted layout of the experimental setup originally shown in Fig.~1 of Ref.~\cite{Smaract}. Panel (a) shows the arrangement of the sensor, mirror and actuator, as well as the effective springs formed by the foam blocks and rubber pads. Panel (b) is a photo of the setup inside the accoustically isolating box.}
    \label{fig:layout}
\end{figure}

We experimentally demonstrate the nonlinear degradation of sensitivity by driving the measurement mirror with a known, high-RMS signal and observing the sensor's resulting spectrum. We make use of a moving magnet actuator, specifically a BOSEM shadow sensor~\cite{Bosem} with the sensing components and circuitry removed. The coil resistance is \SI{41.4}{\ohm} with an inductance of \SI{17.8}{\mH}. The layout can be found in Fig.~\ref{fig:layout}. The sensing scheme and physical layout is identical to the scheme in Ref.~\cite{Smaract} and shown in Fig.~1 therein, except for the additions specified below.

We must isolate the sensor from environmental disturbances, such as seismic and acoustic couplings, as these generate uncontrolled and dominant sources of noise. To achieve this, we set up the mirror-sensor system in an acoustically isolated box on a common base plate, which is placed on foam blocks for further vibration isolation. As the sensor only measures the relative displacement between itself and the mirror, the vibrational coupling to the readout can be suppressed by many orders of magnitude. In a departure from the setup in Ref.~\cite{Smaract}, we isolate the measurement mirror from the common base plate with a rubber pad (stiffer than the foam blocks) to allow some compliance to differential motion. We actuate with the coil on a stack of three RS Pro neodymium magnets (stock number 219-2231) attached to the mirror base, which allows for some residual differential drive, although most of the driven displacement remains common.

We use a ThorLabs LDC 205 C laser diode driver to provide the coil drive current. We monitor this drive current using the built-in control port of the current driver. This is essential as we find that the current driver cannot naturally drive inductance linearly, which we verified by comparing the linearity of the coil drive against the linearity of driving an equivalent resistor. This actuator nonlinearity dominates over the nonlinearity of the sensor and hence must be suppressed. We notice an even stronger nonlinearity when using a mirror-mounted PZT, which leads us to discount this otherwise much simpler actuation scheme. To improve the actuator linearity, we design and implement a feedback system where we subtract the measured drive current from the desired setpoint drive to form an error signal that controls the current driver in-loop. This scheme suppresses the sub-kilohertz nonlinear noise by around two orders of magnitude compared to the free-running performance.

We drive the current with uniform white noise, which is then band-limited with an eighth-order elliptical bandpass filter to produce a large flat signal spectrum confined to a sharply cut off frequency window. With this scheme, we can achieve a displacement RMS of around \SI{0.2}{\um} and three orders of magnitude higher spectral density in the signal window than the residual noise at frequencies below the signal band. The RMS is sufficiently low that the displacement does not significantly change the modulation index (around 1~ppm) and so does not introduce a further source of nonlinearity.

We drive the signal injection digitally and shape this drive using digital filters using the same CDS architecture as is used to read out the sensor. Therefore, we can freely shift the amplitude and frequency band of the signal to investigate the different levels of nonlinear noise that appear at low frequencies. We also simulate the sensor response in the time domain to compare the measured noise with the nonlinear noise models derived in Sec.~\ref{sec:model}.

\begin{figure}[H]
    \centering
    \begin{subfigure}{\linewidth}
        \centering
        \caption{Measured}
        \vspace{2mm}
        \includegraphics[width=\linewidth]{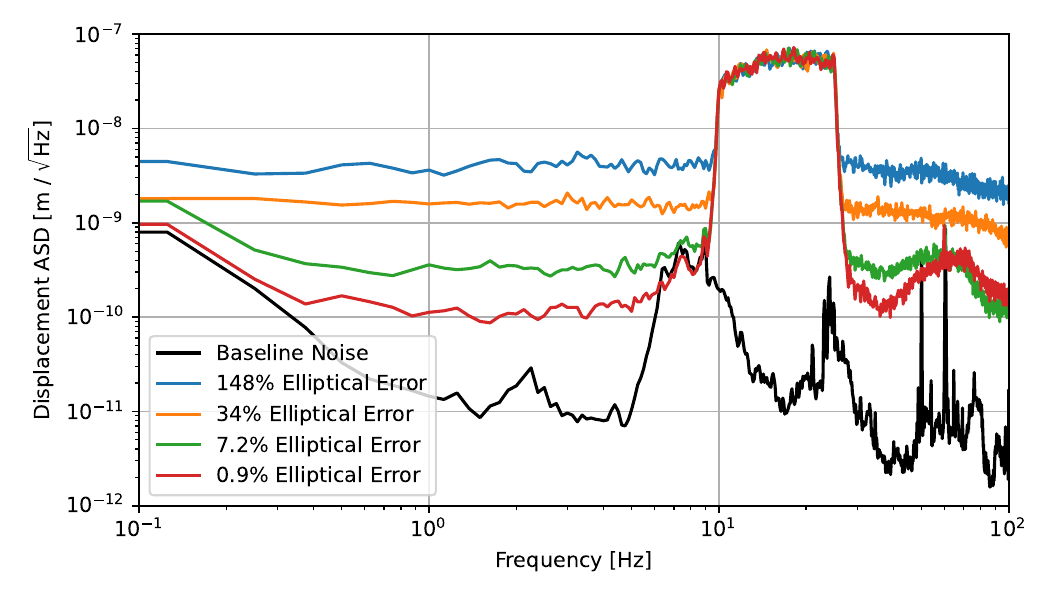}
        \label{subfig:ellip_meas}
    \end{subfigure}
    \hfill
    \begin{subfigure}{\linewidth}
        \centering
        \caption{Simulated}
        \vspace{2mm}
        \includegraphics[width=\linewidth]{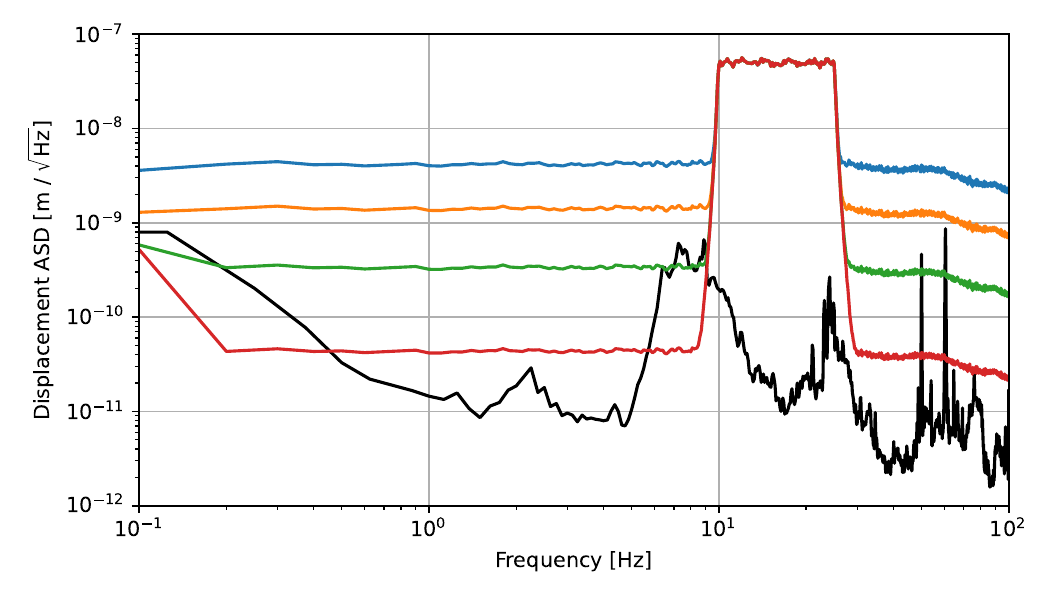}
        \label{subfig:ellip_sim}
    \end{subfigure}
    \caption{Comparison between measured and simulated nonlinearities during a controlled and tightly band-limited injection of displacement noise using a coil-magnet actuator. The simulated spectra show good agreement with measurement for high values of elliptical error, which supports analysis in Sec.~\ref{subsec:ellip}. At the smallest value of elliptical error, the measured nonlinear noise is a factor of a few above the simulation, which can be attributed to the presence of other nonlinearities, particularly the nonelliptical nonlinearity laid out in Sec.~\ref{subsec:nonellip}. The baseline noise curve shows the sum of all measured noises in the quiet (`zero-displacement') statea and represents the absolute noise floor of the sensor.}
    \label{fig:ellip}
\end{figure}

Figure~\ref{fig:ellip} shows the result of deliberately inducing nonlinearities through varying ellipticity. The nonlinear noise is matched well by our time-domain simulation, particularly for large values of the elliptical error, $\delta$. The difference between the measured and simulated noise for the smallest $\delta$ can be explained through the presence of other nonlinearities, which begin to dominate at small values of $\delta$. Whilst some of this discrepancy can be attributed to the residual nonlinearity of the drive, it can be entirely explained by the limiting nonelliptical nonlinearity discussed in Sec.~\ref{subsec:nonellip}, which we show to have a similar effective contribution as a $\delta$ of 2\%. In this scheme we are operating sufficiently below the velocity limit of imposed in Sec.~\ref{subsec:demod}, which means this source of nonlinearity should not limit the readout.

\section{Sensitivity Projections in the LIGO Vacuum Chambers}
\label{sec:LIGO}

We have demonstrated the loss of sensitivity due to nonlinearities in a carefully contrived scenario. However, any scenario which requires access to the full proposed dynamic range of the sensor may quite possibly encounter problems with the sensor linearity. We mentioned the application in inertial sensors, which is highly relevant to the field of GW detection. An even closer application is in the sensing of the multi-stage suspension systems that are found in all contemporary GW detectors. In this section we take the example of the Advanced LIGO quadruple pendulum suspensions~\cite{LIGOQUAD}.

The suspensions in the Advanced LIGO chambers are already placed within a seismically isolated environment on top of the so-called ISI. The suspension stages provide progressively greater levels of vibration filtering, which means their motion is, over many frequencies, even smaller than the residual motion injected by the ISI. Therefore, this application is intuitively less susceptible to nonlinearities due to its low RMS displacement.

We consider, specifically, the measurement of longitudinal displacement sensing of the top mass of the quadruple chain. This is the location of the sensing and control of most of the 24 suspension degrees of freedom and the location of the majority of the existing displacement sensors. We propose our device as a candidate for replacing precisely these sensors in order to achieve the required factor of 100 improvement in the suspension sensing noise, as laid out in Ref.~\cite{5Hz}.

\begin{figure}[H]
    \centering
    \includegraphics[width=\linewidth]{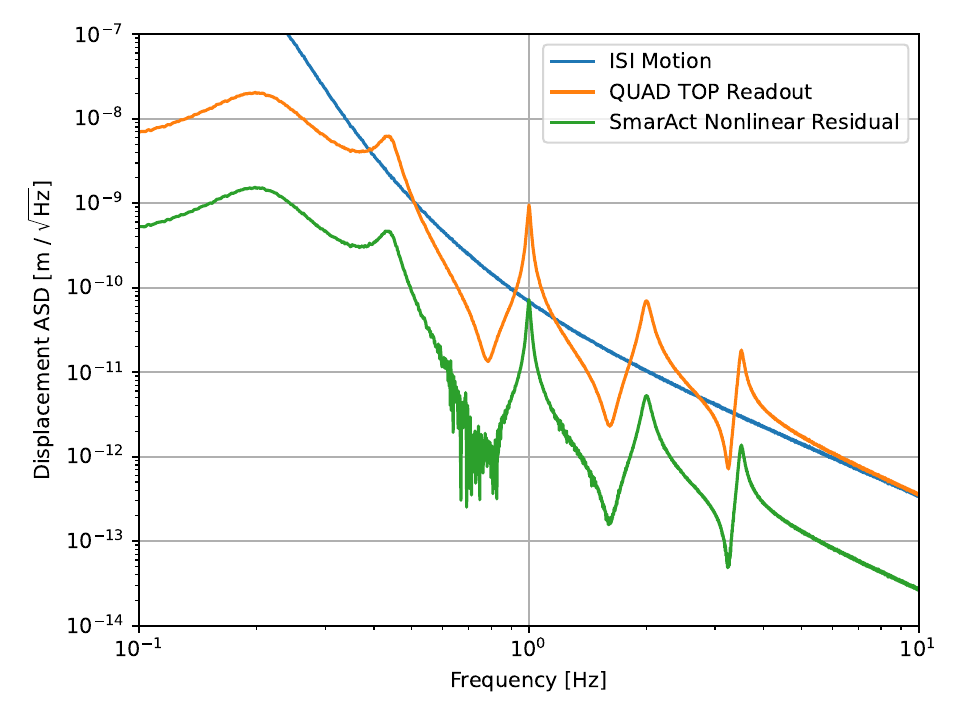}
    \caption{Simulated readout of the top stage of the LIGO quadruple suspension with the SmarAct sensor together with the corresponding residual nonlinear error. Whilst the residual shows nonlinear noise significantly exceeds the nominal noise floor of the device, at no point in the spectrum does the nonlinear residual exceed the measured signal.}
    \label{fig:quad_sim}
\end{figure}

We generate an ISI noise model based on the measured ISI displacement spectral density. We further introduce a fit to the ISI-horizontal-to-top-mass-horizontal transfer function in order to estimate the representative relative displacement spectrum between the top mass and ISI. We pass this displacement spectrum through our model of the sensor nonlinearity, assuming a nominal elliptical error of 2\%. As found in our investigations above, it is possible to reach hard limit of the nonelliptical nonlinearities at this level. The simulated displacement spectrum with the corresponding nonlinear noise level is shown in Fig.~\ref{fig:quad_sim}. As shown, the nonlinear noise can be significantly higher than the quoted sensitivity during a null measurement (Fig. 2b in Ref.~\cite{Smaract}). However, the spectral density of the nonlinearity clearly shadows the spectral density of the signal at a level that is around a factor of 10 lower across the whole frequency band of interest.

\section{Conclusion}
\label{sec:conc}

Interferometric displacement sensors are well poised to replace many existing electromagnetic and optical sensing schemes within applications requiring sub-picometre levels of sensitivity. In this paper we follow up the investigation of our custom sensor manufactured by SmarAct from Ref.~\cite{Smaract}, with a focus on the sensor's linearity. We embed this investigation within the specific context of inertial sensing and gravitational wave detection. The former use case is motivated by the sensor's future implementation on the compact six-degree-of-freedom inertial sensor prototyped in Ref.~\cite{C6D}. The latter is in recognition of the sensor's suitablity as a future candidate sensor on the upgraded Advanced LIGO quadruple suspensions~\cite{LIGOQUAD}.

We briefly lay out the key equations that describe the deep frequency modulation technique that we employ within our readout scheme. From this starting point, we show the possible origins of the nonlinear couplings to the sensor readout and analyse their impact on the displacement sensitivity. We find that an imprecise fit (or drift) of the ellipticity of the Lissajous figure constructed from the two orthogonal readout signals is often the dominant source of nonlinear noise for elliptical error in excess of 2\%. We subsequently revisit our measurement of the current hard limit to the nonlinearity, which is generated by periodic error due to distortion of the Lissajous figure, which cannot be corrected in real-time or in post-processing. This leads to an effective elliptical error of 2\%.

We construct a band-limited, high-RMS scenario in which the nonlinear error can significantly exceed both the linear noise level and the true displacement spectrum at low frequencies. We conduct an experimental demonstration of this nonlinear noise and compare the results to a time-domain simulation, which show good agreement with each other. We also estimate an order-of-magnitude figure for the maximum nonlinear noise level in the presence of a broadband signal and find that it is consistent with measurement.

Overall, the linearity of the sensor is something that should be carefully considered based on the precise parameters of a given application, particularly the RMS displacement and the required dynamic range. We find that the nonlinear noise may limit the sensitivity of inertial sensors if not managed well. However, we find that on relatively quiet platforms, such as the Advanced LIGO ISI, the linearity of the sensor is sufficient to ensure an SNR above unity within the detection band, and thus no significant improvements to the sensor performance are necessary.

Despite the generally sufficient linearity of the sensor for applications that at the heart of our investigation, there is certainly scope for further improvement of the sensor's linearity. Future experimental work on the sensor we presented herein will focus on implementing more advanced algorithms that seek to improve long-term stability and, potentially, real-time correction and linearisation of the system.

\authorcontributions{Conceptualization, D.M.; methodology, J.S. and D.M.; investigation, J.S., C.D.F., A.A. and D.M.; resources, D.M.; data curation, J.S. and D.M.; writing---original draft preparation, J.S.; writing---review and editing, C.D.F., A.A. and D.M.; visualization, J.S.; supervision, D.M.; project administration, D.M.; funding acquisition, D.M. All authors have read and agreed to the published version of the manuscript.}

\funding{This research was funded by STFC grant numbers ST/T006609/1 and ST/W006375/1 and EPSRC grant numbers EP/V008617/1.}

\acknowledgments{We thank members of the LIGO Suspension Working Group for useful discussions. J.S. and D. M. acknowledge the support of the Institute for Gravitational Wave Astronomy at the University of Birmingham, STFC Quantum Technology for Fundamental Physics schemes (Grant No. ST/T006609/1 and ST/W006375/1), and EPSRC New Investigator Award (Grant No. EP/V008617/1).  D.M. is supported by the 2021 Philip Leverhulme Prize.
}

\conflictsofinterest{The authors declare no conflict of interest. The funders had no role in the design of the study; in the collection, analyses, or interpretation of data; in the writing of the manuscript; or in the decision to publish the~results.}

\abbreviations{Abbreviations}{
The following abbreviations are used in this manuscript:\\

\noindent 
\begin{tabular}{@{}ll}
BOSEM & Birmingham Optical Sensor and Electromagnetic Motor\\
BRS & Beam Rotation Sensor\\
DFM & Deep frequency modulation\\
DoF & Degree of freedom\\
FSR & Free spectral range\\
GW & Gravitational wave\\
HoQI & Homodyne Quadrature Interferometer\\
ISI & Internal Seismic Isolation\\
LIGO & Laser Interferometer Gravitational-Wave Observatory\\
LISA & Laser Interferometer Space Antenna\\
PSD & Power spectral density\\
PZT & Piezoelectric transducer\\
RMS & Root mean square\\
SNR & Signal-to-noise ratio
\end{tabular}
}

\begin{adjustwidth}{-\extralength}{0cm}

\reftitle{References}
\bibliography{nonlin}


\PublishersNote{}
\end{adjustwidth}
\end{document}